\documentclass[aps, prd,floats, floatfix, twocolumn,superscriptaddress, nofootinbib, showpacs]{revtex4-1}

\usepackage{graphicx}
\usepackage{hyperref}
\usepackage{amsmath,amssymb}
\usepackage{amsfonts}
\usepackage{xspace} % Sensible space treatment at end of simple macros
\usepackage[usenames]{color}
\usepackage{dcolumn} % Align table columns on decimal point
\usepackage{bm} % bold math
\usepackage{epsfig}

\usepackage{ulem}
\definecolor {darkgreen}{rgb}{0.2,0.7,0.2}

\newcommand\be{\begin{equation}}
\newcommand\ba{\begin{eqnarray}}
\newcommand\ee{\end{equation}}
\newcommand\ea{\end{eqnarray}}
\newcommand{\Sec}{{\mbox{\tiny Sec}}}
\newcommand{\Tot}{{\mbox{\tiny Tot}}}
\newcommand{\tidal}{{\mbox{\tiny Tidal}}}
\newcommand{\orb}{{\mbox{\tiny Orb}}}
\newcommand{\los}{{\mbox{\tiny los}}}
\newcommand{\Newt}{{\mbox{\tiny Newt}}}

\newcommand{\GW}{{\mbox{\tiny GW}}}

\newcommand{\Acc}{{\mbox{\tiny Acc}}}
\newcommand{\noAcc}{{\mbox{\tiny no.~Acc}}}

\newcommand{\ISCO}{{\mbox{\tiny ISCO}}}

\newcommand{\gta}{\mathrel{\raise.3ex\hbox{$>$}\mkern-14mu
             \lower0.6ex\hbox{$\sim$}}}
\newcommand{\lta}{\mathrel{\raise.3ex\hbox{$<$}\mkern-14mu
             \lower0.6ex\hbox{$\sim$}}}
\newcommand{\SMBH}{\bullet}
\newcommand{\SCO}{\star}

\begin{document}

\title{The Effect of Massive Perturbers on Extreme Mass-Ratio Inspiral Waveforms}

\author{Nicol\'as Yunes}
\affiliation{Department of Physics and MIT Kavli Institute, 77 Massachusetts Avenue, Cambridge, MA 02139, USA.}
\affiliation{Harvard-Smithsonian, Center for Astrophysics, 60 Garden St., Cambridge, MA 02138, USA.}

\author{M.~Coleman Miller}
\affiliation{Maryland Astronomy Center for Theory and Computation \& Joint Space-Science Institute, 
 Department of Astronomy, University of Maryland, College Park, MD 20742, USA.}

\author{Jonathan Thornburg}
\affiliation{Department of Astronomy and Center for Spacetime Symmetries, Indiana University, Bloomington, IN 47405, USA.}

%-----------------------------------------------------------------------------
\begin{abstract}
Extreme mass ratio inspirals, in which a stellar-mass object orbits a supermassive black hole, 
are prime sources for space-based gravitational wave detectors because
they will facilitate tests of strong gravity and probe the
spacetime around rotating compact objects.  In the last few
years of such inspirals, the total phase is in the millions of
radians and details of the waveforms are sensitive to
small perturbations. We show that one potentially detectable
perturbation is the presence of a second supermassive black hole
within a few tenths of a parsec. The acceleration produced by the
perturber on the extreme mass-ratio system produces a steady drift 
that causes the waveform to deviate systematically from that of an isolated
system.  If the perturber is a few tenths of a parsec from the
extreme-mass ratio system (plausible in as many as a few percent of cases) 
higher derivatives of motion might also be detectable. 
In that case, the mass and distance of the perturber can be
derived independently, which would allow a new probe of merger
dynamics.
\end{abstract}

\date{\today}
\maketitle

%-----------------------------------------------------------------------------
\section{Introduction}
\label{sec:intro}

Space-based gravitational wave detectors such as the {\it Laser
Interferometer Space Antenna} ({\it LISA}) are expected to see
a wide variety of sources in their $\sim 10^{-4}-10^{-1}$~Hz
sensitivity band.  Of these, extreme mass-ratio inspirals (EMRIs), 
a stellar-mass compact object (SCO) spiraling 
into a supermassive black hole (SMBH),
are considered particularly promising because they can probe 
strong gravity over millions of radians of phase evolution
in the last few years of evolution.
As a result, EMRI waveforms serve
as highly precise probes of strong gravity and of the spacetime
around rotating SMBHs.  Considerable study has been devoted to
astrophysical scenarios for EMRIs
~\cite{2005ApJ...631L.117M,Levin:2006uc,2007CQGra..24..113A} 
as well as to the analysis of their waveforms
~\cite{Hughes:1999bq,Hughes:2001jr,Barack:2003fp,Babak:2006uv,Yunes:2009ef,2009GWN.....2....3Y,2010arXiv1009.6013Y}.

There has been less exploration of the possibility of deviations
from isolated EMRI waveforms that might occur due to environmental
effects (see e.g.~\cite{Giampieri:1993pt,Chakrabarti:1995dw,2000ApJ...536..663N,Levin:2006uc,2008PhRvD..77j4027B} for a study 
of differences caused by an accretion disk around the SMBH).  
Here we point out an effect that has not been considered in this context: the acceleration of the
EMRI system by a nearby (distance of roughly a few tenths of a parsec or less)
secondary SMBH.  As we demonstrate, this acceleration
leads to phase drifts of fractions of a radian over a year of
inspiral, which is potentially detectable from EMRIs of plausible
signal strength.  Depending on the fraction of galaxies that 
merge, and on the fraction of time in such mergers that the
secondary SMBH is within a few tenths of a parsec of the primary, this could
affect as many as a few percent of EMRIs.  

The detection of such an effect could yield a new probe of 
galactic merger dynamics, providing a measure of the ratio of the secondary
SMBH's mass and its distance to the EMRI. If such effects are not present in 
a detected gravitational wave (GW), then one can place an upper limit
on the density of SMBHs inside some radius of a few tenths of a parsec.  
If this is the case, then one would confirm that, as far as LISA is concerned, 
EMRIs occur in vacuum. 

This paper is organized as follows:
In Sec.~\ref{sec:simple} we do a simple analysis of the acceleration effect as
it would apply to a signal of constant frequency and amplitude, 
which we expand on in the Appendix.
In Sec.~\ref{sec:realistic} we explain how to model real EMRI waveforms, for 
the particular case of quasi-circular, equatorial orbits, and explain
how to implement modifications to model an acceleration effect.
In Sec.~\ref{sec:pert-effect} we extend the simple analysis of Sec.~\ref{sec:simple} to real waveforms
and perform a dephasing and an overlap study. 
In Sec.~\ref{sec:degeneracies} we explore whether some of these deviations can be 
masked by adjustments of EMRI system parameters.  
We present our conclusions in Sec.~\ref{sec:discussion}. In most of this paper, 
we use geometric units with $G = c = 1$. For reference, in this system
of units, one solar mass 
$M_{\odot} = 1.476 \; {\rm{km}} = 4.92 \times 10^{-6} \; {\rm{s}}$, 
while 
$1 \; {\rm{pc}} = 1.03 \times 10^{8} \; {\rm{s}} = 2.09 \times 10^{13} M_{\odot}$.

%-----------------------------------------------------------------------------
\section{Simple Model}
\label{sec:simple}

\begin{figure}[t]
\includegraphics[width=8.5cm,clip=true]{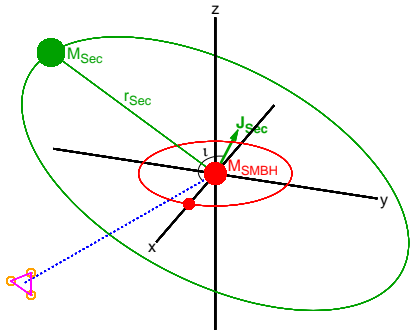} 
 \caption{\label{fig:sketch} Schematic view of the EMRI system (in the $xy$~plane), the massive
  perturber $M_\Sec$ (at a distance $r_\Sec$ from the EMRI SMBH),
  and the line of sight.  $\iota$ is the inclination between the
  primary-secondary SMBH's orbital angular momentum vector and the
  line of sight.}
\end{figure}

Here we present the basic effects of acceleration in a simplified
model.  We assume that there is an EMRI of a SCO into 
a (primary) SMBH with mass $M_{\SMBH}$ on the $\hat{x}$-$\hat{y}$ plane, 
with orbital and spin angular momentum in the $\hat{z}$ direction. We further
simplify the scenario by assuming GWs of constant
frequency and amplitude. Let us also assume there is a secondary SMBH in a circular
orbit about the EMRI's center of mass (COM).  Suppose that the secondary SMBH has a mass 
$M_{\Sec}$ and the total mass of the system $M_{\Tot} = M_{\SMBH}+M_{\Sec}$. 
Suppose also that the semi-major axis of the circular orbit of the primary-secondary SMBH system 
is $r_{\Sec}$, and that it is inclined to the line of sight at an angle $\iota$ 
(here $\iota$ is zero for a face-on binary and $90^\circ$ for an edge-on binary).  
We depict this scenario in Fig.~\ref{fig:sketch}. 

If these systems are well-separated, then the EMRI's COM will
move essentially at a constant velocity relative to us, with a 
projection into our line of sight of
\begin{equation}
v_{\los}(t)=\left( \frac{M_{\Sec}}{M_{\Tot}}\right)
v_{\Newt} \cos\left(\omega_{\Newt} t + \delta \right) \; \sin\left(\iota\right)\;.
\label{vlos}
\end{equation}
where $v_{\Newt} = (G M_{\Tot}/r_{\Sec})^{1/2}$ is the Newtonian virial velocity, 
$\omega_{\Newt} = (G M_{\Tot}/r_{\Sec}^{3})^{1/2}$ is the Newtonian angular velocity
for an object in a circular orbit, and $\delta$ is an initial phase, with $(\omega_{\Newt} t + \delta)$
the orbital phase of the EMRI-Secondary system.
A constant relative speed is entirely absorbed in a 
redefinition of the masses. As such, constant relative velocities 
cannot enter any of our results.  

If the EMRI's COM is sufficiently close to the secondary SMBH, then the former
will experience an measurable {\emph{acceleration}},  which will produce
a net Doppler phase drift relative to the best-fit waveform.  
Note that the orbital period for $M_{\Tot}=10^{6-7}~M_\odot$
and $r_{\Sec}=0.1-1$~pc is at least $\sim 10^{3-5}$~years, so for the
duration of a LISA observation the binary will not change
phase significantly. Tidal effects on the EMRI system due to
the perturber can be neglected, as
this acceleration scales as the inverse cube of $r_{\Sec}$
(see Sec.~\ref{sec:mod-EOB} for more details). 

If the EMRI system is accelerated by an amount ${\dot v}$
relative to its original line of sight speed over a time
$t$, then the GW phase difference compared to the initial frequency
is $\Delta\Phi_{\GW}={1\over 2}{\dot v}tN/c$, where $N$ is the 
number of radians in the waveform (see Appendix~\ref{app:acc-effect} 
for a more detailed explanation of this effect).  Let us designate by
$\epsilon$ the detectable fractional phase shift:
$\epsilon\equiv\Delta\Phi_{\GW, detect}/N$.  As a fiducial
value we will use $\epsilon=10^{-7}$, or 0.1 radians over
$\sim 10^6$ radians for a typical one-year inspiral.
To leading-order in a Taylor-expansion about $\omega_{\Newt}t = 0$, 
we then have
\begin{equation}
{1\over 2} \frac{{\dot v} t}{c}={1\over 2} \left(\sin\iota\right) \left(\sin\delta\right)
{M_{\Sec}\over{r_{\Sec}}} \frac{t}{r_{\Sec}} =\epsilon\;,
\label{eq-cons1}
\end{equation}
where we have here neglected a constant term that is non-observable. 
Solving for the distance at which this is satisfied gives
\begin{eqnarray}
r_{\Sec} &\approx& 0.26~{\rm pc} \; \left(\sin\iota\right)^{1/2} \left(\sin\delta\right)^{1/2}
\left(\frac{M_{\Sec}}{10^6~M_\odot}\right)^{1/2}
\nonumber \\
&\times&
\left(\frac{t}{1~{\rm yr}}\right)^{1/2}
\left(\frac{\epsilon}{10^{-7}}\right)^{-1/2}\; .
\end{eqnarray}

The next order term in the phase shift scales as
\begin{equation}
\frac{1}{6} \frac{{\ddot v}t^2}{c} = {1\over 6} \left(\sin\iota \right) \left(\cos\delta\right)
\frac{M_{\Sec}}{r_{\Sec}}
\sqrt{\frac{M_{\Tot}}{r_{\Sec}}} \frac{t^2}{r_{\Sec}^{2}}\,.
\label{eq-cons2}
\end{equation}
Setting this equal to $\epsilon$ and solving for $r$, we find
\begin{eqnarray}
r_{\Sec} &\approx& 0.025~{\rm pc}\;\left(\sin\iota\right)^{2/7} \left(\cos\delta\right)^{2/7}
\left(\frac{M_{\Sec}}{10^6~M_\odot}\right)^{2/7}
\nonumber \\
&\times&
\left(\frac{M_{\Tot}}{2 \times 10^6~M_\odot}\right)^{1/7}
\left(\frac{t}{1~{\rm yr}}\right)^{4/7}
\left(\frac{\epsilon}{10^{-7}}\right)^{-2/7}\; .
\end{eqnarray}
Additional corrections can be computed similarly.

Therefore, for BH masses $\gta 10^6~M_\odot$ and separations
of a few tenths of a parsec or less, acceleration can cause
a detectable shift in the simplified waveform. As we find in 
Sec.~\ref{sec:pert-effect}, this shift is proportional
to the combination ${\cal{A}} \equiv M_{\Sec}/r_{\Sec}^{2}$. 
For separations of a few hundredths of a parsec or less, higher order derivatives
are measurable. In this case, the detectable shift in the waveform is captured
by the linear combination of ${\cal{A}}$ and other higher-order derivative terms,
such as ${\cal{B}} \equiv M_{\Sec}^{3/2} r_{\Sec}^{-7/2}$. Given a sufficiently 
small $r_{\Sec}$ one could then measure both ${\cal{A}}$ and ${\cal{B}}$ and
thus disentangle $M_{\Sec}$ from $r_{\Sec}$.

\begin{figure}[t]
\includegraphics[width=8.5cm,clip=true]{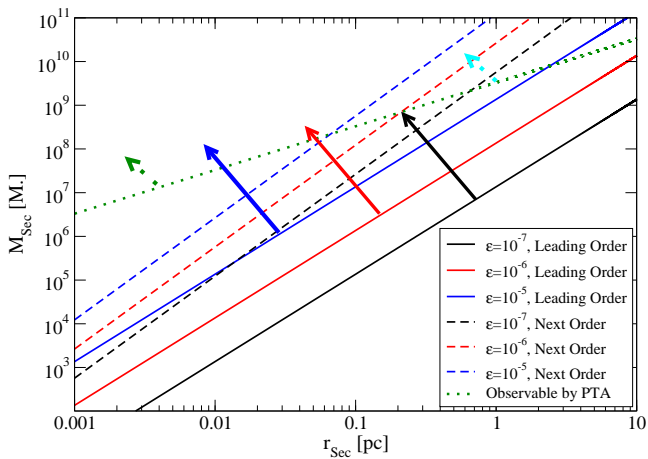} 
 \caption{\label{fig:FOM} Range of secondary masses and separations that could
 be measurable by LISA given a sufficiently strong EMRI. The region above the solid and dashed
 lines would be observable.  Measurement of the leading-order effect gives a determination
  of the combination $(M_\Sec \sin\iota)/r_\Sec^2$, while measuring
  the next-order effect gives a determination of the combination
  $(M_\Sec \sin\iota)^{3/2}/r_\Sec^{7/2}$.  Thus measuring both
  effects together allows both $M_\Sec \sin\iota$ and $r_\Sec$ to
  be determined.}
\end{figure}

The range of masses and separations that could be observed, given a sufficiently
strong EMRI-perturber system are depicted in Fig.~\ref{fig:FOM}. In this figure, we show
with solid lines the constraint given by Eq.~\eqref{eq-cons1}, and with a dashed line
that of Eq.~\eqref{eq-cons2} (with $M_{\SMBH} = M_{\Sec}$ for simplicity), where the black, 
red and blue colors correspond to $\epsilon = 10^{-7}$, $10^{-6}$ and $10^{-5}$.  A larger
value of $\epsilon$ corresponds to more conservative choices of what is detectable by LISA. 
The area above the curves show the values of $M_{\Sec}$ and $r_{\Sec}$ that could be measurable. 
For comparison, we also show the region of $(M_{\Sec},r_{\Sec})$ space that fall in the pulsar-timing-array
(PTA) sensitivity band. Of course, for PTAs to individually resolve such binaries, their distance to Earth
would have to be sufficiently small~\cite{2010CQGra..27h4013H}. 
In principle, however, this scenario allows for the possibility of
coincident future detection of GWs with LISA and PTAs.

We now discuss the detectability of the changes discussed above for a realistic EMRI. 
In Sec.~\ref{sec:discussion}, we return to the question of how common it will be to have a secondary SMBH this close.

%-----------------------------------------------------------------------------
\section{Realistic EMRI Waveforms}
\label{sec:realistic}

%-----------------------------------------------------------------------------
\subsection{Standard EOB Modeling}
\label{standard-EOB}

We employ the effective-one-body (EOB) formalism to model waveforms
with and without the acceleration correction. This formalism was initially developed in~\cite{Buonanno99,Buonanno00} to model comparable-mass BH binary coalescences.
Improvements and extensions to other binaries were developed in~\cite{Damour00,Damour01,Buonanno06,Damour:2008qf,Barausse:2009xi,Nagar:2006xv,Damour2007,Bernuzzi:2010ty,Damour:1997ub,Damour2007,Damour:2008gu,Pan:2010hz,Fujita:2010xj} and compared to a set of numerical relativity results in~\cite{Damour:2009kr,Buonanno:2009qa,Pan:2009wj} and to self-force calculations in~\cite{Barack:2009ey,Damour:2009sm}. Recently,~\cite{Yunes:2009ef,2009GWN.....2....3Y,2010arXiv1009.6013Y} combined the EOB approach with BH perturbation theory results to model EMRI waveforms for LISA data-analysis. We here concentrate on the formulation of~\cite{2010arXiv1009.6013Y}, as it is applicable to EMRIs, the systems of interest in this paper. 

We focus on quasi-circular EMRI inspirals in the equatorial plane of a spinning BH because they are simpler to model. We define the following orbital parameters: the 
SCO's mass $m_{\SCO}$; the SMBH's mass $M_{\SMBH}$; the total mass $M = m_{\SCO}
+ M_{\SMBH}$; the reduced mass $\mu = m_{\SCO} m_{\SMBH}/M$; and the symmetric mass-ratio $\nu = \mu/M$. We also assume that the EMRI's orbital angular momentum is aligned with the MBH's spin angular momentum $S_{\SMBH} = a_{\SMBH} M_{\SMBH} = q_{\SMBH} M_{\SMBH}^2$, where $a_{\SMBH}=S_{\SMBH}/M_{\SMBH}$ is the MBH's spin parameter and $q_{\SMBH} = a_{\SMBH}/M_{\SMBH}$ is its dimensionless spin parameter. We employ the adiabatic approximation, in which we assume that the radiation-reaction time-scale is much longer than the orbital one. 

With this at hand, let us now describe the EOB approach we employ. In the adiabatic approximation, the GW phase can be obtained by solving
\begin{eqnarray}
\label{omegaadiab}
\dot{\omega} &=& - \left(\frac{d E}{d \omega}\right)^{-1} \, {\cal F}(\omega)\,,
\\
\dot{\phi} &=& \omega\,, 
\label{phiadiab}
\end{eqnarray}
where $\omega \equiv \dot{\phi}$ is the orbital angular frequency, with $\phi$ the orbital phase, overhead dots stand for time derivatives, $E$ is the system's total energy and ${\cal F}$ is the GW energy flux. The energy of the system is~\cite{Bardeen:1972fi}
\begin{equation}
E=M_{\SMBH} + m_{\SCO}\,\frac{1-2 M_{\SMBH}/r \pm q_{\SMBH}\,M_{\SMBH}^{3/2}/r^{3/2}}{\sqrt{1-3 M_{\SMBH}/r\pm 2q_{\SMBH}\,M_{\SMBH}^{3/2}/r^{3/2}}}\,.
\label{E-binding}
\end{equation}
where the $\pm$ stands for prograde or retrograde orbits. In this equation and all throughout the rest of this paper, we ignore sub-leading corrections that are proportional to the EMRI's mass-ratio. In practice, this means we ignore conservative and second-order dissipative self-force effects, i.e.~the effect of the SCO on its own geometry, as well as the SCO's spin.  

The GW flux can be written in the factorized form of~\cite{Damour2007,Damour:2008gu,Pan:2010hz,2010arXiv1009.6013Y}, which in the adiabatic regime is
\begin{eqnarray}
{\cal F}(\omega)= \frac{1}{8 \pi}\,\sum_{\ell=2}^{8}\sum_{m=0}^{\ell} (m\,\omega)^2\,\left|h_{\ell m} \right|^{2}\,,
\label{pd-flux-adiab}
\end{eqnarray}
where the multipole-decomposed waveforms are 
\begin{equation}
h_{\ell m}(v) = h_{\ell m}^{\Newt,\epsilon_p}\;S^{\epsilon_p}_{\ell m} \; T_{\ell m}\; e^{i \delta_{\ell m}}\; (\rho_{\ell m})^\ell\,,
\label{full-h}
\end{equation}
and where $\epsilon_p$ is the parity of the waveform (\textit{i.e.,} $\epsilon_p=0$ if $\ell+ m$ is even, $\epsilon_p=1$ if $\ell+ m$ is odd). The quantities ($S^{\epsilon_p}_{\ell m}(v)$, $T_{\ell m}(v)$, $\delta_{\ell m}(v)$ and $\rho_{\ell m}(v)$) in Eq.~\eqref{full-h} can be found in~\cite{Damour2007,Damour:2008gu,Pan:2010hz}.  The Newtonian waveform is 
\begin{equation}
h_{\ell m}^{\Newt,\epsilon_p} \equiv \frac{M_{\SMBH}}{R}\, n^{(\epsilon_p)}_{\ell m}\, c_{\ell +\epsilon_p}\, v^{\ell +\epsilon_p}\, 
Y_{\ell - \epsilon_p,-m}(\pi/2,\phi).
\end{equation}
where $Y_{\ell,m}(\theta,\phi)$ are spherical harmonic functions, while $n_{\ell m}^{(\epsilon_p)}$ and $c_{\ell +\epsilon_p}$ are numerical coefficients~\cite{Damour:2008gu}. 

We enhance the flux of Eq.~\eqref{pd-flux-adiab} by linearly adding BH absorption terms and calibration coefficients that are fitted to a more accurate, numerical flux~\cite{2010arXiv1009.6013Y}. The first modification is necessary as BHs lose energy due to GWs that both escape to infinity and fall into BHs. The second modification accounts for the fact that the bare fluxes written above are built from low-velocity (PN) expansions, and as such, are not sufficiently accurate by themselves for long evolutions, even after the resummations introduced.

The above differential system is solved with the post-circular initial conditions of~\cite{Buonanno00}, enhanced with a mock-evolution at $100 M_{\SMBH}$ (see e.~g.~\cite{2010arXiv1009.6013Y}). The orbital phase can then be used in the waveforms of Eq.~\eqref{full-h}, together with the fact that for quasi-circular orbits
\begin{equation}
r = \frac{\left[1 - q_{1}\, (M_{\SMBH} \omega) \right]^{2/3}}{(M_{\SMBH} \omega)^{2/3}}\,.
\label{r-of-w}
\end{equation}
where $r$ is the EMRI's separation and $v = (M_{\SMBH} \omega)^{1/3}$ by Kepler's third law. With the waves at hand, we then compute the GW phase and its amplitude via
\begin{equation}
\Phi_{\GW}^{\ell m} = \Im \left[ \ln \left(\frac{h_{\ell m}}{|h_{\ell m}|} \right)\right]\,, \qquad
A_{\GW}^{\ell m} = |h_{\ell m}|\,.
\end{equation}
The GW phase as defined above needs to be unwrapped every $2 \pi$, so in
practice it is simpler to define the time derivative of this quantity and then
obtain $\Phi_{\GW}^{\ell m}$ via integration.

%-----------------------------------------------------------------------------
\subsection{Modifications to EOB Modeling}
\label{sec:mod-EOB}

How do we incorporate the effects of an external acceleration into GW modeling within the EOB framework? Let us first distinguish between wave generation and wave propagation effects. By the former, we mean effects that arise in the {\emph{near-zone}} (less than a gravitational wavelength away from the EMRI's COM) and that generate GWs due to the inspiral of the EMRI. By the latter, we mean effects that arise after the system has generated a GW and it propagates out to the {\emph{wave-zone}}, where the observer is located, many gravitational wavelengths away from the source. 

As is expected, all propagation effects, such as the backscattering (or tails) of GWs off the metric of the secondary SMBH, occur beyond Newtonian (leading) order in post-Newtonian theory~\cite{Blanchet:2002av}, and can be safely neglected here. The presence of an external source, however, does introduce non-negligible modifications to the generation of GWs. One could incorporate such effects by introducing an external, vectorial force to Hamilton's equations in the direction of the perturber. This force would simply be the product of the the total mass of the system and the time derivative of the velocity of Eq.~\eqref{vlos}. The modeling of this effect would require a non-adiabatic evolution, i.e.~the evolution of the full set of Hamilton's equations, without assuming circular orbits or using Kepler's third law. One expects that such force would induce eccentricity and inclination in the inspiral, driving the SCO out of the equatorial plane of the secondary SMBH. 

One can estimate the magnitude of this effect by considering the tidal force effect of the perturber on the COM relative to the SCO's acceleration due to the secondary SMBH. 
Since the tidal force scales as $F_{\tidal} = M_{\Sec}/r_{\Sec}^{3}$, this effect is suppressed relative to the acceleration by a factor of $r/r_{\Sec} \sim 10^{-4}$ for
an EMRI with orbital separation of $30 m_{\SMBH}$ and a primary-secondary SMBH orbital separation of $0.01$ pc.  The ratio is this small because the perturber is assumed to be at parsec scales away from the COM, and one parsec translates to $\sim 10^{13} M_{\odot}$ in geometric units. Since the tidal force scales as the inverse of the separation cubed, any tidal effects are insignificant.  

Given that this type of generation effects are suppressed, are there any others
that should be included? The dominant generation effect is simply a Doppler shift in the frequencies, which then leads to an integrated modification in the GW phase (see the Appendix for a detailed explanation of this Doppler effect). In this sense, such a correction is similar to the integrated Sachs-Wolfe effect for GWs~\cite{2010ApJ...715L..12L}, where here the perturbation to the potential is given by a third body, instead of some cosmological background. The implementation of this correction to an EOB evolution is simple: divide the right-hand-side of Eq.~\eqref{phiadiab} by the appropriate Doppler factor
\begin{equation}
\dot{\phi} = \omega \quad \rightarrow \quad \dot{\phi} = \omega \; \left[1 + v_{\los}(t,\delta=\pi/2) \right]\,.
\end{equation}
In this equation, we have not included the appropriate Lorentz factor $\Gamma$, since $v_{\Newt}/c \ll1$, and we can linearize in this quantity. Moreover, we have removed the constant velocity drift component of $v_{\los}$ by choosing $\delta = \pi/2$, as the former is not measurable.

%-----------------------------------------------------------------------------
\section{Perturbing Acceleration Effect on Relativistic EMRI Waveforms}
\label{sec:pert-effect}

%--------------------------------
\subsection{Preliminary Considerations} 
With the machinery described in Sec.~\ref{sec:realistic}, we can 
construct modified EMRI waveforms as a function of time, for a given 
value of the second MBH mass and separation to the EMRI's COM. We
consider the following two EMRI systems, integrated for one year each:
\begin{itemize}
\item {\bf{System I}}: The primary SMBH has mass $m_{\SMBH} = 10^{5} \,
  M_{\odot}$ and spin parameter $q_{\SMBH} = 0.9$, while the SCO has mass 
  and spin parameter $m_{\SCO} = 10 \, M_{\odot}$ and $q_{\SCO} = 0$.
  This system inspirals for $\sim 6 \times 10^{5}$ rads
  of orbital phase between orbital separations
  $r/M \in (16,26)$.  In this range the orbital velocities are
  $v \in (0.2,0.25)$ and the GW frequencies are
  $f_{\GW} \in (0.005,0.01) \; {\rm{Hz}}$. 
\item {\bf{System II}}: The primary SMBH has mass $m_{\SMBH} = 10^{6} \,
  M_{\odot}$ and spin parameter $q_{\SMBH} = 0.9$, while the SCO has mass 
  and spin parameter $m_{\SCO} = 10 \, M_{\odot}$ and $q_{\SCO} = 0$.
  This system inspirals for $\sim 3 \times 10^{5}$ rads
  of orbital phase between orbital separations
  $r/M \in (11,r_{\ISCO})$.  In this range the orbital velocities are
  $v \in (0.3,v_{\ISCO})$ and the GW frequencies are
  $f_{\GW} \in (0.001,f_{\GW}^{\ISCO}) \; {\rm{Hz}}$. 
\end{itemize}
System~I exits the most sensitive part of the LISA band at around 
$16 M$, which is why we stop the evolution there. In contrast, Sys.~II 
is stopped when the SCO reaches the innermost stable circular orbit (ISCO). 
For each of these systems, we explore a variety of secondary SMBH 
masses $M_{\Sec} = (10^{5}, 10^{6}) M_{\odot}$ as well as a 
variety of separations $r_{\Sec} = (0.01,0.1,1) \; {\rm{pc}}$. 
Larger secondary masses are also possible; these would have equivalent
effects on the EMRI at correspondingly larger distances $r \sim M^{1/2}$.
(For example, $M_{\Sec} = 10^9 M_{\odot}$ at $r = 30$~pc would have
equivalent effects to $M_{\Sec} = 10^6 M_{\odot}$ at $r = 1$~pc.) All of this 
information is summarized in Table~\ref{system-summary}, including the orbital
periods $T_{\orb}$ and the time to merger due to GW emission $T_{\GW}$.
In all cases we set $\sin{\iota} = 1$ and $\delta = \pi/2$, as this leads to 
the largest possible effect. The reasoning behind this is that if this effect is not
observable with this choice of parameters, it will not be observable with any other
choice.

\begin{table}
\begin{ruledtabular}
\begin{tabular}{cccccccc}
$m_{\SMBH}$ & $m_{\SCO}$ & $M_{\Sec}$ & $r_{\Sec}/{\rm{pc}}$ 
& $T_{\orb}^{\SCO}$ & $T_{\orb}^{\SMBH}$ & $T_{\GW}^{\SCO}$ & $T_{\GW}^{\SMBH}$ \\
$10$ & $10^{5}$ &$10^{5}$ &$10^{-2}$ &$10^{-5}$ &$2.1 \times 10^{2}$ &$5.6$ &$10^{19}$ \\
$10$ & $10^{5}$ &$10^{5}$ &$10^{-1}$ &$10^{-5}$ &$6.6 \times 10^{3}$ &$5.6$ &$10^{23}$ \\
$10$ & $10^{5}$ &$10^{5}$ &$10^{+0}$ &$10^{-5}$ &$2.1 \times 10^{5}$ &$5.6$ &$10^{27}$ \\
\hline
$10$ & $10^{5}$ &$10^{6}$ &$10^{-2}$ &$10^{-5}$ &$8.9 \times 10^{1}$ &$5.6$ &$10^{18}$ \\
$10$ & $10^{5}$ &$10^{6}$ &$10^{-1}$ &$10^{-5}$ &$2.8 \times 10^{3}$ &$5.6$ &$10^{22}$ \\
$10$ & $10^{5}$ &$10^{6}$ &$10^{+0}$ &$10^{-5}$ &$8.9 \times 10^{4}$ &$5.6$ &$10^{26}$ \\
\hline
\hline
$10$ & $10^{6}$ &$10^{5}$ &$10^{-2}$ &$10^{-5}$ &$8.9 \times 10^{2}$ &$18$ &$10^{18}$ \\
$10$ & $10^{6}$ &$10^{5}$ &$10^{-1}$ &$10^{-5}$ &$2.8 \times 10^{3}$ &$18$ &$10^{22}$ \\
$10$ & $10^{6}$ &$10^{5}$ &$10^{+0}$ &$10^{-5}$ &$8.9 \times 10^{4}$ &$18$ &$10^{26}$ \\
\hline
$10$ & $10^{6}$ &$10^{6}$ &$10^{-2}$ &$10^{-5}$ &$6.6 \times 10^{2}$ &$18$ &$10^{18}$ \\
$10$ & $10^{6}$ &$10^{6}$ &$10^{-1}$ &$10^{-5}$ &$2.1 \times 10^{3}$ &$18$ &$10^{22}$ \\
$10$ & $10^{6}$ &$10^{6}$ &$10^{+0}$ &$10^{-5}$ &$6.6 \times 10^{4}$ &$18$ &$10^{26}$ \\
\end{tabular}
\caption{\label{system-summary}Summary of System properties. All masses are in units of $M_{\odot}$,
$r_{\Sec}$ is in units of parsecs and all time scales are in units of years. The time to merger is here estimated
as $T_{\GW} = r/\dot{r}_{\GW}$, where $\dot{r}_{\GW}$ is the rate of change of the orbital separation due to 
GW emission and for $r$ we take the values in the itemized list. The superscript star stands for quantities
associated with the SCO-SMBH system, while the solid dot stands for those associated with the SMBH-SMBH
system.}
\end{ruledtabular}
\end{table}

%--------------------------------
\subsection{Dephasing Study} 

Let us define the dephasing between waveforms as follows:
\begin{equation}
\Delta \Phi_{\GW} \equiv \Phi_{\GW}^{\Acc} - \Phi_{\GW}^{\noAcc}\,
\end{equation}
where $\Phi_{\GW}^{\Acc}$ is the GW phase of an EMRI waveform with 
an accelerated COM, while $\Phi_{\GW}^{\noAcc}$ is that of an inertial
COM. We have here aligned the waveforms in time and phase before
computing this dephasing. This alignment is equivalent to minimizing the 
statistic in Eq.~(23) of~\cite{Buonanno:2009qa}, which in turn is the same
as maximizing the fitting factor over time and phase of coalescence in a matched
filtering calculation with white noise~\cite{Buonanno:2009qa}. The
alignment is done here in the same way as in~\cite{Yunes:2009ef,2009GWN.....2....3Y,2010arXiv1009.6013Y}.

\begin{figure}[t]
\includegraphics[width=8.5cm,clip=true]{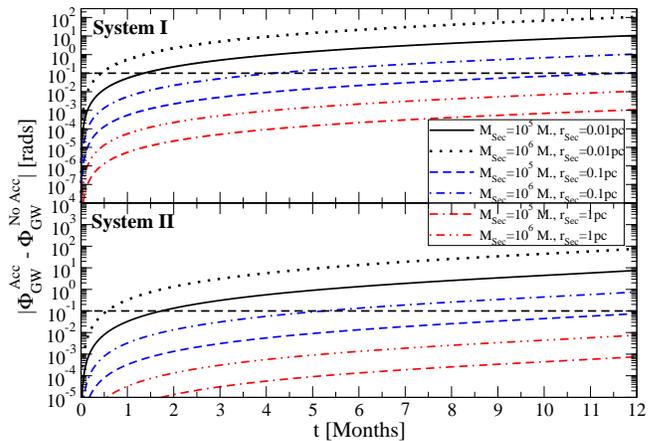} 
 \caption{\label{fig:dephasing} Dephasing for Sys.~I and II as a function of time
 in units of months, for a variety of separations and masses of the perturber.}
\end{figure}

Figure~\ref{fig:dephasing} plots the dephasing of the dominant $(\ell,m) = (2,2)$ 
GW mode as a function of time in months for Sys.~I and II. 
The different line colors/shades correspond to different separations to the perturber [$r_{\Sec} = (0.01, 0.1, 1)$ pc], while different line styles correspond
to different perturber masses [$M_{\Sec} = (10^{5},10^{6}) M_{\odot}$].  Observe that for both 
Systems a dephasing of order $0.1$ rads is achieved for separations $r_{\Sec} \lesssim 0.1$ pc 
over less than one year. This is consistent with the estimates of Sec.~\ref{sec:simple}. Similarly, 
more massive perturbers enhance the dephasing roughly by one order of magnitude. 
The amplitudes of the waveforms are not shown in this figure; they disagree at the level of $10^{-3}$ for Sys.~I and $10^{-4}$ for Sys.~II.  

The magnitude and shape of the dephasing depends on how far away and massive the perturber is. One can show that the dephasing scales as $\Delta \Phi_{\GW}  \propto N M_{\Sec} T/r_{\Sec}^{2}$, where $T$ is the observation time and $N$ is the number of cycles. Since there is a factor of two fewer GW cycles in Sys.~II relative to Sys.~I, then the dephasing for the former is also smaller by a factor of two.

%--------------------------------
\subsection{Overlap Study} 

The dephasing study of the previous subsection is suggestive, but not sufficiently quantitative to assess whether such types of corrections are large enough to be measurable. Let us then perform a slightly more sophisticated data analysis study here. 

Given any time series $A(t)$ and $B(t)$, one can construct the inner-product
\begin{equation}
\left( A \right| \left. B \right) = 4\, {\rm Re} \int_{0}^{\infty} \frac{\tilde{A}(f) \; \tilde{B}^{\star}(f)}{S_{n}(f)} \, df\,
\end{equation}
where the overhead tildes stand for the Fourier transform, the star stands for
complex conjugation and $S_{n}(f)$ is the spectral density of noise in the detector. We choose here the sky-averaged version of the noise curve presented in~\cite{2004PhRvD..70l2002B,2005PhRvD..71h4025B}. 

With this inner-product, we can now construct some data analysis measures. The 
signal to noise ratio (SNR) of signal $A$ is
\begin{equation}
\rho(A) = \sqrt{\left( A \right| \left. A \right)}\,,
\end{equation}
while the overlap between signals $a$ and $b$ is
\begin{equation}
{\rm M} = {\rm{max}} \frac{\left( A \right| \left. B \right)}{\sqrt{{\left( A \right| \left. A \right) {\left( B \right| \left. B \right)}}}}\,.
\label{match-def}
\end{equation}
with the mismatch ${\rm MM} = 1 - {\rm M}$. The max label in Eq.~\eqref{match-def} is to remind us that this statistic must be maximized over an event time (e.g., the time of coalescence of the EMRI system) and a phase shift~\cite{Damour:1997ub}. If the overlap is larger than $97 \%$ (or equivalently, if the mismatch is lower than $3 \%$), then the difference between waveforms $A$ and $B$ is sufficiently small to not matter for detection purposes (see e.~g.~\cite{Owen:1995tm}). The minimum overlap quoted above ($97\%)$ is mostly conventional, motivated by the fact that the event rate scales as the cube of the overlap for a reasonable source distribution. For an overlap larger than $97 \%$, no more than $10 \%$ of events are expected to be lost at SNRs of ${\cal{O}}(10)$. Of course, for larger SNRs, one might not need such high overlaps, although EMRI sources are expected to have SNRs $< 100$. 

Whether the difference between waveforms $A$ and $B$ can be detected in parameter estimation can be assessed by computing the SNR of the difference in the waveforms $\delta h\equiv A - B$:
\begin{equation}
\rho(\delta h) = \sqrt{\left( \delta h\right| \left. \delta h\right)}
=  4\, {\rm Re} \int_{0}^{\infty} \frac{\tilde{\delta h}(f) \; \tilde{\delta h}^{\star}(f)}{S_{n}(f)} \, df\,.
\end{equation}
When this SNR equals unity, then one can claim that $A$ and $B$ are sufficiently dissimilar that they can be differentiated via matched filtering (see e.~g.~\cite{2010arXiv1008.1803L}).

\begin{figure}[t]
\includegraphics[width=8.5cm,clip=true]{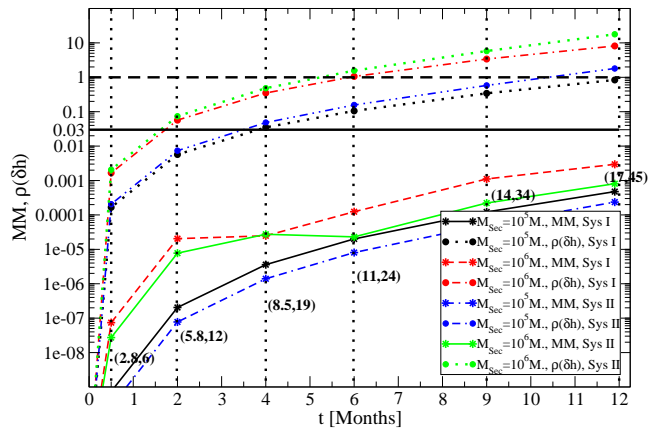} 
 \caption{\label{fig:overlap} Mismatch as a function of time in units of months for Sys.~I and II and different perturber masses, all at a separation of $r_{\Sec} = 0.1 \; {\rm{pc}}$. 
 SNRs for Sys.~I and~II are given in parentheses for a source at $1$ Gpc.}
\end{figure}

We applied these measures to EOB waveforms with and without acceleration of the COM. The results are plotted in Fig.~\ref{fig:overlap} as a function of observation time in months. The vertical dotted lines correspond to observation times of $(0.5, 2, 4, 6, 9, 12)$ months, and the numbers next to them, in parenthesis, stand for the SNR of Sys.~I and II for that observation time. The different line styles and colors correspond to mismatches and SNRs of the error for different secondary systems. Observe that the mismatch is always smaller than $0.03$ (the solid black horizontal line), suggesting that this effect will not affect detection. Observe also that the SNR of the difference reaches unity (the dashed black horizontal line) in between 6 and 12 months of observation, and for the $M_{\rm Sec}=10^{6}~M_{\odot}$, $\rho(\delta{\rm h})$ reaches $\sim 10$ after one year. This suggests that given a sufficiently strong EMRI with ${\rm SNR}\sim 50-100$, the magnitude of this effect is in principle detectable within one year of coherent integration.
%

%-----------------------------------------------------------------------------
\section{Degeneracies}
\label{sec:degeneracies}

Now that we have determined that there exists a set of plausible perturber
parameters for which the magnitude of the correction could be measurable, let us
consider the possibility of degeneracies. That is, let us investigate whether 
we can mimic an acceleration of the COM by
changing the intrinsic parameters (the component masses, the spin parameter, etc.)
in the non-accelerating waveform. The simplest way to see whether this is possible is to 
study the frequency dependence of the GW modification introduced by the 
COM's acceleration. 

Let us then remind ourselves of how the frequency-domain representation is constructed. For this, we employ the stationary-phase approximation (see e.~g.~\cite{2009PhRvD..80h4001Y}), under which, the frequency-domain waveform is simply
\begin{equation}
\tilde{h}(f) = {\cal{A}} f^{-7/6} e^{i \psi(f)}\,,
\end{equation}
where the Newtonian (leading-order) amplitude is ${\cal{A}} = \pi^{-2/3} {30}^{-1/2} \; {\cal{M}}^{5/6} D_{L}^{-1}$, with ${\cal{M}} = \eta^{3/5} M$, while the phase is constructed from
\begin{equation}
\psi(f) = - \frac{\pi}{4} + 2 \pi f t(f) - 2 \phi(f)\,,
\label{phase-SPA}
\end{equation}
where the second term arises due to the Fourier transform and the third term due to the oscillatory nature of the time-domain waveform. 

The phase of the frequency-domain waveform in the stationary phase 
approximation is then controlled by these last two terms in 
Eq.~\eqref{phase-SPA}. The first term can be computed via
\begin{equation}
2 \pi f t(f) = 2 \pi f \int^{f/2} \frac{\tau(F^\prime)}{F^\prime} dF^\prime\,,
\end{equation}
where $f$ is the GW frequency, while the second term can be calculated from 
\begin{equation}
\phi(f) = 2 \pi \int^{f/2} \tau(F^\prime) dF^\prime\,,
\end{equation}
where $\tau(F) \equiv F/\dot{F}$ and $F$ is the orbital frequency. 

The Doppler correction to the waveform comes in the calculation of $\phi(f)$, as this is simply the integral of the frequency. For simplicity, we can reparameterize the $v_{\los}(t) \to v_{\los}(F)$, by noting that, to Newtonian order, 
\begin{equation}
2 \pi F(t) = \frac{4^{-3/2}}{M} \left[ \frac{\eta}{5 M} \left(t_{c} - t \right) \right]^{-3/8}\,, 
\label{Foft}
\end{equation}
which we can invert to obtain
\begin{equation}
t(F) = t_{c} - \frac{5 M}{256 \eta} \left( 2 \pi M F\right)^{-8/3}\,,
\end{equation}
where $t_{c}$ is the time of coalescence of the EMRI system. Taylor-expanding Eq.~\eqref{vlos} about $\omega_{\Newt} t = 0$, we find  
\begin{eqnarray}
v_{\los} &\sim& v_{0} + v_{1} \left(2 \pi M F\right)^{-8/3} 
+ v_{2} \left(2 \pi M F\right)^{-16/3} 
\nonumber \\
&+& v_{3} \left(2 \pi M F\right)^{-8}\,,
\end{eqnarray}
where the $v_{i}$ coefficient are the
following frequency-independent functions:
\begin{eqnarray}
v_{0} &=& \left(\frac{M_{\Sec}}{M_{\Tot}} \right)^{1/2} 
\left(\frac{M_{\Sec}}{r_{\Sec}}\right)^{1/2}
\left[ \cos\delta - 
\left(\frac{M_{\Tot}}{r_{\Sec}}\right)^{1/2} \frac{t_{c}}{r_{\Sec}} \sin{\delta} 
\right.
\nonumber \\
&-& \left.
 \frac{1}{2} \frac{M_{\Tot} t_{c}^2}{r_{\Sec}^{3}} \cos\delta 
+ \frac{1}{2} \left(\frac{M_{\Tot}}{r_{\Sec}}\right)^{3/2} \frac{t_{c}^3}{r_{\Sec}^{3}} \sin\delta
 \right] \sin{\iota}
\nonumber \\
v_{1} &=&
\frac{10}{512} \frac{M M_{\Sec}}{r_{\Sec}^{2}} \eta^{-1} (\sin{\iota}) 
\left[ \sin\delta + 
\left(\frac{M_{\Tot}}{r_{\Sec}} \right)^{1/2} \frac{t_{c}}{r_{\Sec}} \cos\delta 
\right. 
\nonumber \\
&-& \left.
- \frac{1}{2} \frac{M_{\Tot} t_{c}^2}{r_{\Sec}^{3}} \sin{\delta} \right]
\nonumber \\
v_{2} &=& \frac{25}{131072} \frac{M^2 M_{\Sec} }{r_{\Sec}^{3}}
\left[-\left(\frac{M_{\Tot}}{r_{\Sec}}\right)^{1/2}  \cos{\delta} 
+ \frac{M_{\Tot} t_{c}}{r_{\Sec}^{2}} \sin{\delta}\right]
\nonumber \\
&\times& 
\eta^{-2} \sin{\iota} \,, 
\nonumber \\
v_{3} &=& - \frac{125}{100663296} \frac{M^3 M_{\Sec} M_{\Tot}}{r_{\Sec}^5} \eta^{-3}  \sin{\delta} \sin{\iota}\,.
\end{eqnarray}
Notice that $v_{0}$ is of ${\cal{O}}(M^{1/2}/r_{\Sec}^{1/2})$, $v_{1}$ is of ${\cal{O}}(M^{2}/r_{\Sec}^{2})$, $v_{2}$ is of ${\cal{O}}(M^{7/2}/r_{\Sec}^{7/2})$ and $v_{3}$ is of ${\cal{O}}(M^{5}/r_{\Sec}^{5})$.

With these relations at hand, we can now compute the correction to the frequency-domain waveform phase in the stationary phase approximation. Denoting by $\Delta \psi = \psi_{\Acc - }\psi_{\noAcc}$, we find that
\begin{eqnarray}
\Delta \psi &=& -4 \pi \int^{f/2} \tau(F') \; v_{\los}(F') \; dF'\,,
\nonumber \\
&\sim&- \frac{5 \pi}{24} \frac{M}{\eta} \int^{f/2} \left(2 \pi M F'\right)^{-8/3} 
\left[v_{0} + 
v_{1} \left(2 \pi M F'\right)^{-8/3} 
\right. 
\nonumber \\
&+& \left. v_{2} \left(2 \pi M F'\right)^{-16} +
v_{3} \left(2 \pi M F'\right)^{-8} \right] dF'
\end{eqnarray}
where in the second line we have used that to Newtonian order 
\begin{equation}
\dot{F} = \frac{48}{5 \pi {\cal{M}}^{2}} \left(2 \pi {\cal{M}} F \right)^{11/3}\,.
\label{dotF}
\end{equation}
Normalizing this phase correction by the Newtonian form of the frequency-domain waveform phase, we find
\begin{eqnarray}
\Delta \psi &=& \frac{3}{128} \left(\pi {\cal{M}} f\right)^{-5/3} \left[ 
\frac{8}{3} v_{0}  
+  \frac{40}{39} v_{1} \eta^{8/5} \left(\pi {\cal{M}} f\right)^{-8/3}
\right. 
\nonumber \\
&+& \left.
 \frac{40}{63} v_{2} \eta^{16/5} \left(\pi {\cal{M}} f\right)^{-16/3}
+  \frac{40}{39} v_{3} \eta^{24/5} \left(\pi {\cal{M}} f\right)^{-8}
\right]\,.
\label{SPA-correction}
\end{eqnarray}
Setting $\delta = \pi/2 = \iota$, the above expression simplifies to
\begin{eqnarray}
\Delta \psi &=& \frac{3}{128} \left(\pi {\cal{M}} f\right)^{-5/3} \left[ 
-\frac{8}{3} \frac{M_{\Sec} t_{c}}{r_{\Sec}^{2}} 
+ \frac{4}{9} \frac{M_{\Sec} M_{\Tot} t_{c}^{2}}{r_{\Sec}^{5}}
\right. 
\nonumber \\
&+&\left. \left(\frac{25}{1248} \frac{{\cal{M}} M_{\Sec}}{r_{\Sec}^{2}}
- \frac{25}{2496} \frac{{\cal{M}} M_{\Sec} M_{\Tot} t_{c}^{2}}{r_{\Sec}^{5}}
\right) \left(\pi {\cal{M}} f\right)^{-8/3}
\right. 
\nonumber \\
&+&\left.
\frac{125}{1032192} \frac{{\cal{M}}^{2} M_{\Sec} M_{\Tot} t_{c}}{r_{\Sec}^{5}}
\left(\pi {\cal{M}} f\right)^{-16/3}
\right. 
\nonumber \\
&-&\left.
\frac{625}{1094713344} \frac{{\cal{M}}^{3} M_{\Sec} M_{\Tot}}{r_{\Sec}^{5}}
\left(\pi {\cal{M}} f\right)^{-8}
\right]\,.
\label{SPA-correction2}
\end{eqnarray}

Let us now discuss this result in more detail. The first two terms inside the square bracket in Eq.~\eqref{SPA-correction2} arise due to a constant misalignment between the time of coalescence of the EMRI system and the primary-secondary SMBH system (we have implicitly set the latter to zero). This effect can be absorbed via a redefinition of the chirp mass, and thus, it is not observable. All other lines in Eq.~\eqref{SPA-correction2}, on the other hand, contain a non-trivial frequency-dependence and they cannot be reabsorbed via a redefinition of intrinsic parameters. 

A physical way to think about this is the following. Given a signal and a template without modeling a secondary perturber, one would like to maximize 
the phase coherence by shifting the template's phase and frequency. Such a shift corresponds to an adjustment of the total mass and the chirp mass, which eliminates 
the first term in Eq.~\eqref{SPA-correction2}. Once this shift is done, however, there are no other template parameters that can be shifted, while the frequency 
derivatives of the signal and template will continue to disagree. 

The many terms that arise in the second, third and fourth lines of Eq.~\eqref{SPA-correction2} are due to the Taylor expansion in $\omega_{\Newt} t$,   which in frequency space has become an expansion in inverse powers of $(M \, f)$ and $(r_{\Sec}/M_{\Sec})$. Clearly, the second line Eq.~\eqref{SPA-correction2} is dominant over all others as it scales with $r_{\Sec}^{-2}$ to leading order, while the third and fourth lines scale as $r_{\Sec}^{-5}$. If $M_{\Sec}/r_{\Sec}$ is large enough, however, one might be able to measure the coefficients in front of both the dominant $f^{-8/3}$ term and the $f^{-16/3}$ or $f^{-8}$ term. This would then imply that one could break the degeneracy between $M_{\Sec}$ and $r_{\Sec}$ in the leading order term and measure both quantities. 

Ignoring such possible acceleration effects could introduce a bias in the extraction of parameters via matched filtering~\cite{Cutler:2007mi,2009PhRvD..80l2003Y}. Imagine, for example, that an EMRI GW is detected with vacuum templates. One would then proceed to extract parameters from this detection, such as the primary SMBH's and SCO's mass with some error bars. Usually, the error estimate accounts for statistical error plus possible systematics with the modeling. The acceleration effect here would be one such systematic, whose magnitude would have to be determined via a careful Markov-Chain Monte-Carlo exploration of the likelihood surface. 

Notice also that in Eq.~\eqref{SPA-correction2} we have kept only the Newtonian contribution to an infinite post-Newtonian expansion. This is essentially because in Eqs.~\eqref{Foft} and~\eqref{dotF} we have dropped all but the leading order, Newtonian term. Interestingly, the correction terms that arise at leading order are dominant over the Newtonian piece, as they depend on high inverse powers of frequency (in particular, higher than $5/3$). This implies that if a detailed parameter
estimation study were to be carried out, these post-Newtonian terms should be 
taken into account, as they contribute at the same order as the Newtonian
term in an inertial frame.

The dependence of the correction in Eq.~\eqref{SPA-correction2} on different powers of the frequency suggests that these terms are non-degenerate with the standard ones that appear in the non-accelerating GW phase. More precisely, the GW phase in an inertial frame is given by (see eg.~\cite{2009PhRvD..80h4001Y}). 
\begin{eqnarray}
\psi(f)_{\noAcc} &=& 2 \pi f t_{c} - \phi_{c} + \frac{3}{128} \left(\pi {\cal{M}} f\right)^{-5/3}
\\ \nonumber 
&\times&
 \left[ 1 + \left( \frac{3715}{756} + \frac{55}{9} \eta \right) \eta^{-2/5} \left( \pi {\cal{M}} f \right)^{2/3} + \ldots \right]\,,
\end{eqnarray}
where $\phi_{c}$ is the phase of coalescence and the ellipses stand for higher order terms in the post-Newtonian series. Notice that there are no powers of $f^{-8/3}$, $f^{-16/3}$ or $f^{-8}$ in the above equation. Thus, the correction computed in Eq.~\eqref{SPA-correction2} is weakly correlated to the GW phase in an inertial frame, i.~e.~the off-diagonal elements of the Fisher matrix are small for the $M_{\Sec}/r_{\Sec}^{2}$ coordinate sector relative to the diagonal term. Although these results are suggestive, a more detailed analysis should be carried out to determine the level of correlation between all parameters and the accuracy to which $M_{\Sec}$ and $r_{\Sec}$ could be extracted.

Although the correction due to the acceleration of the COM seems to be weakly
correlated to other intrinsic parameters, one might
wonder whether it is degenerate with other effects not included in vacuum GR waveforms. Takahashi and Nakamura~\cite{2005PThPh.113...63T} have studied the effect of the acceleration of the Universe in the frequency-domain form of the waveform. They find that
\begin{equation}
\Delta \psi = \frac{3}{128} \left(\pi {\cal{M}} f\right)^{-5/3} \left[\frac{25}{768} {\cal{M}} \dot{z} \left( \pi {\cal{M}}  f \right)^{-8/3} \right]\,.
\label{cosmo-effect}
\end{equation}
One can clearly see that this cosmological effect is degenerate with the one computed here [the second line in Eq.~\eqref{SPA-correction2}]. However, the magnitude of Eq.~\eqref{cosmo-effect} is much smaller than that of Eq.~\eqref{SPA-correction2}, simply because $H_{0} \ll M_{\Sec}/r_{\Sec}^{2}$ for all relevant perturbers considered here. For example, at small redshift, $H_{0} \sim 10^{-23} \; {\rm{km}}^{-1}$ in geometric units, while at $r_{\Sec} = 0.1$ pc and for a $10^{6} M_{\odot}$ perturber, $M_{\Sec}/r_{\Sec}^{2} \sim 10^{-19} \; {\rm{km}}^{-1}$. The perturber separation at which these effects become comparable is approximately $r_{\Sec} \sim 11 \; {\rm{pc}} \; [M_{\sec}/(10^{6} M_{\odot})]^{1/2}$. 

Another possible source of degeneracy could be if there are corrections to general relativity that induce phase modifications with the specific frequency dependence found in Eq.~\eqref{SPA-correction2}. In fact, we see that the result obtained here can be mapped to the parameterized post-Einsteinian framework~\cite{2009PhRvD..80l2003Y} with the choice
\begin{eqnarray}
\alpha = 0, \qquad
\beta = \frac{25}{1248} \frac{M_{\Sec} {\cal{M}}}{r_{\Sec}}, \qquad
b = -\frac{8}{3}\,,
\end{eqnarray}
to leading order in $M_{\Sec}/r_{\Sec}$ (see e.g.~Eq.~$1$ in~\cite{2009PhRvD..80l2003Y}). As found in that paper, however, there are no known alternative theories to date that could potentially lead to the frequency dependence found in Eq.~\eqref{SPA-correction2}. 

%-----------------------------------------------------------------------------
\section{Discussion and Conclusions}
\label{sec:discussion}

We have shown that a $\sim 10^6 M_{\odot}$ secondary SMBH within a few tenths of
a parsec of the EMRI system can produce detectable modifications
in the waveform. A more massive secondary SMBH at a correspondingly larger
distance would produce equivalent effects. It is not possible to say with certainty how common
this will be.  A rough upper limit can be obtained from the 
following observation. Since a redshift of $z=1$ (corresponding roughly to
$10^{10}$ years), tens of percent of Milky Way-like galaxies have
had a major merger~\cite{2006ApJ...652..270B,2008ApJ...672..177L}.  If the typical merger takes hundreds
of millions of years, then at most a few percent will be involved
in a merger at any stage.  The fraction of time spent at
separations $\lta 1$~pc remains uncertain; although
there are well-understood dynamical processes that can reduce
the secondary SMBH's separation to $\sim 1$~pc and gravitational
radiation will bring the binary to merger from $\sim 10^{-3}$~pc, 
the transition between the regimes is uncertain
(this is commonly called the ``final parsec problem"; see, e.g.,
\cite{2003ApJ...596..860M}
for a discussion).  It is therefore possible that the system
spends considerable time at roughly the detectable separations.

We also note that when a secondary SMBH comes within a few tenths
of a parsec of the primary, various dynamical effects temporarily
increase the rate of close encounters of stellar-mass objects with 
both SMBHs~\cite{2009ApJ...697L.149C}.  As a result, it may be that  a 
disproportionate number of EMRIs occur with a secondary SMBH nearby.
Indeed, recently~\cite{Chen:2010wy} estimated that more than $10\%$ of all tidal
disruption events could originate in massive black hole binaries,
so if the EMRI fraction is similar it corresponds to our rough
estimate.

In these cases, measurement of an EMRI phase shift affords a new
way to detect the presence of a binary SMBH.  If the separation is
close enough to measure an additional derivative of the motion,
then the degeneracy between the secondary mass and its distance
is broken. If the EMRI-SMBH system is sufficiently close, then pulsar
timing measurements~\cite{2010CQGra..27h4013H} might
also be able to detect gravitational waves from the SMBH-SMBH binary.
Alternatively, if no phase shift is detected, then this implies that
there are no secondary SMBH in a radius of a few tenths of a parsec, thus
implying an upper limit on the density of BHs close to the detected EMRIs. 
In principle, therefore, EMRIs have another astrophysical
link in addition to their utility in testing general relativity. 

The importance of the astrophysical environment in EMRI GW modeling 
is a double-edged sword. Although on the one hand, one could potentially
extract some astrophysical information, on the other,
these effects could make it difficult to test general relativity~\cite{2009PhRvD..80l2003Y}.
For such tests to be possible, one must have complete control of the waveforms within general
relativity. If the astrophysical environment needs to be included, then the modeling
might be dramatically more difficult. We note here, however, that only a fraction of
EMRIs would experience the astrophysical environment effect discussed here. 
If deviations from general relativity are present, on the other hand,  these should be
present for all EMRIs. Thus, in principle, a statistical analysis would allow us to disentangle
deviations in our waveforms to discern whether they have an astrophysical or
theoretical origin. 

%-----------------------------------------------------------------------------
\acknowledgments
We are grateful to Pau Amaro-Seoane and Ed Porter for organizing the
Astro-GR workshop in Paris, where this idea was conceived. We are also
grateful to Bence Kocsis and to the anonymous referee for many useful comments and suggestions. 
MCM acknowledges support from NASA grant NNX08AH29G. 
NY acknowledges support from NASA through the Einstein Postdoctoral Fellowship Award Number PF0-110080 issued by the Chandra X-ray Observatory Center, which is operated by the Smithsonian Astrophysical Observatory for and on behalf of NASA under contract NAS8-03060.

%----------------------------------------------------------------------------------------
\appendix
\section{Acceleration Effect}
\label{app:acc-effect}

Here we explain in more detail how the Doppler correction to the waveform comes about. Let us consider the effect of an acceleration on the COM position vector. For simplicity we consider the toy-model of a perfect circular orbit with angular velocity $\omega$, whose position vector in the COM can be parameterized as
\begin{equation}
\vec{x} = b \left(\cos{\omega t}, \sin{\omega} t, 0\right)\,,
\end{equation}
where $b$ is the binary's separation and we have erected a Cartesian coordinate system, with the binary in the $\hat{x}$-$\hat{y}$ plane. If an external force is present that causes an acceleration, this in turn will cause a displacement $\vec{x} \to \vec{x}' = \vec{x} + \delta \vec{x}$. Let us parameterize the magnitude of this displacement as $|\delta \vec{x}| = (1/2) \dot{v}_{\los} t^{2}$, which holds to Newtonian order for a uniformly accelerated body. One can then show that the shift in the magnitude of the COM velocity vector is simply
\begin{equation}
|\vec{v}'| = |\vec{v}| + \frac{1}{2} \dot{v}_{\los} t \left(\hat{x} \cdot \delta{\hat{x}} \right)
+ {\cal{O}}(\dot{v}_{\los}^{2} t^{2})\,,
\label{vshifteq}
\end{equation}
where $\dot{v} = \dot{\vec{x}}$ is the unperturbed velocity vector. Notice that there is a factor of $1/2$ here, just as in the estimates of Sec.~\ref{sec:simple}.

Before proceeding, it is useful to concentrate on this velocity shift. Choosing $\delta = \pi/2 = \iota$, one can easily show that
\begin{equation}
v_{\los} \sim \frac{M_{\Sec}}{M_{\Tot}} v_{\Newt} (\omega_{\Newt} t) 
\left[1 - \frac{1}{6} \omega_{\Newt}^{2} t^{2} + {\cal{O}}(\omega_{\Newt}^{4} t^{4}) \right]\,,
\label{vlos-eq}
\end{equation}
upon Taylor expanding about $\omega_{\Newt}t \ll 1$. We can take the time-derivative of $v_{\los}$ and then Taylor-expand again to find:
\begin{equation}
\dot{v}_{\los} t 
\sim  - \frac{M_{\Sec}}{M_{\Tot}} v_{\Newt} \; (\omega _{\Newt}t)
\left[1 - \frac{1}{2} \omega_{\Newt}^{2} t^{2} + {\cal{O}}(\omega_{\Newt}^{4} t^{4}) \right]\,. \quad
\label{vlos-eq2}
\end{equation}
Obviously, this is the same as simply Taylor-expanding $v_{\los}$ to leading order.

One effect of the COM velocity drift is a Doppler shift to the waveform. Special relativity predicts that if a frequency source is moving with velocity $v$ away from the observer at an angle $\theta$, then the frequency observed is
\begin{eqnarray}
\omega' &=& \frac{\omega}{\Gamma} \left(1 + v \cos{\theta}\right)^{-1}\,,  
\label{doppler}
\nonumber \\
&\sim&
\omega \left[1 - v \cos{\theta} + v^{2} \left(\cos^{2}{\theta} - \frac{1}{2} \right) + {\cal{O}}(v^{4}) \right],
\label{omega-eq}
\end{eqnarray}
where $\omega$ is the frequency of the source, $\omega'$ is the frequency the observer detects and $\Gamma = (1 - v^{2})^{-1/2}$ is the usual special relativity factor. 
In the notation of Sec.~\ref{sec:simple}, $v \cos{\theta} = v_{\los}$. 

The Doppler shift effect can also be understood in two additional, complementary ways. 
The LISA response function naturally contains a Doppler shift term in the phase, 
due to LISA's motion about the Solar System barycenter. The Doppler shift discussed 
above is identical to this, but now it is the GW source that moves, as opposed to the detector. 
Similarly, one could consider first the GW phase emitted in the EMRI's center of mass, and then
map this to that observed in the Solar System by shifting the phase's time-dependence 
by $dt$, corresponding to the light travel time along the line of sight between the center of mass
of the EMRI system and the SMBH-SMBH system. From this perspective, the maximum phase
shift that could accumulate in one year is simply the product of the EMRI orbital frequency and
the light-crossing time of the projection of one year of SMBH binary evolution
along the line of sight.

We can then easily integrate Eq.~\eqref{omega-eq}, assuming a constant $\omega$, to recover the $\Delta \Phi_{\GW}$ computed in Sec~\ref{sec:simple}. Setting $\iota = \pi/2 = \delta$, we find
\begin{equation}
\Delta \Phi_{\GW} = - \omega \int v_{\los}(t)  dt\,,
\sim -\frac{N}{2} \left(\frac{M_{\Sec}}{M_{\Tot}} \right) \; v_{\Newt} \left({\omega_{\Newt} T} \right) \,.
\end{equation}
Notice that by choosing $\delta = \pi/2$, there is no leading-order, unobservable constant velocity drift term. In the second line, we have Taylor expanded about $\omega_{\Newt} t = 0$ 
and used that $\Phi_{\GW,\Tot} = N =  \omega T$, where $T$ is the time of integration, 
and that $\Phi_{\GW,\Tot} = N$, where $N$ is the total number of radians in the non-accelerating waveform. Notice that this is the same $\Delta \Phi_{\GW}$ correction described in Sec.~\ref{sec:simple}. 

\bibliographystyle{apsrev}
\bibliography{review}
\end{document}